\documentclass[a4paper,fleqn,usenatbib]{mnras}

\usepackage[T1]{fontenc}
\usepackage{ae,aecompl}

\usepackage{graphicx}	
\usepackage{amsmath}	
\usepackage{amssymb}	
\usepackage{subfigure}

\newcommand{\pcc}{\,{\rm cm}^{-3}}

\newcommand{\kel}{\, {\rm K}}

\newcommand{\msun}{\, {\rm M}_\odot}
\newcommand{\nh}{n_{\rm H}}

\newcommand{\pc}{\, {\rm pc}}

\newcommand{\mh}{m_{\rm H}}

\newcommand{\yr}{\, {\rm yr}}
\newcommand{\myr}{\, {\rm Myr}}

\newcommand{\power}[2]{$#1 (#2)$}

\title[Ambipolar diffusion in cores]{Ambipolar diffusion and the molecular abundances in prestellar cores}

\author[Priestley et al.]{
Felix D. Priestley,$^{1}$
James Wurster$^{2}$
and Serena Viti$^{3}$
\\
$^{1}$School of Physics and Astronomy, Cardiff University, Queens Buildings, The Parade, Cardiff CF24 3AA, UK \\
$^{2}$School of Physics and Astronomy, University of Exeter, Stocker Road, Exeter EX4 4QL, UK\\
$^{3}$Department of Physics and Astronomy, University College London, Gower Street, London WC1E 6BT, UK\\
}

\date{Accepted XXX. Received YYY; in original form ZZZ}

\pubyear{2019}

\begin{document}
\label{firstpage}
\pagerange{\pageref{firstpage}--\pageref{lastpage}}
\maketitle

\begin{abstract}
  We investigate differences in the molecular abundances between magnetically super- and sub-critical prestellar cores, performing three-dimensional non-ideal magnetohydrodynamical (MHD) simulations with varying densities and magnetic field strengths, and post-processing the results with a time-dependent gas-grain chemical code. Most molecular species show significantly more central depletion in subcritical models, due to the longer duration of collapse. However, the directly observable quantities - the molecule to hydrogen column density ratios - are generally too similar for observational data to discriminate between models. The profiles of N$_2$H$^+$ and HCO$^+$ show qualitative differences between supercritical and subcritical models on scales of $0.01 \pc$, which may allow the two cases to be distinguished. However, this requires knowledge of the hydrogen column density, which is not directly measureable, and predicted line intensity profiles from radiative transfer modelling are similar for these molecules. Other commonly observed species, such as HCN and CH$_3$OH, have line intensity profiles which differ more strongly between models, and so are more promising as tracers of the mechanism of cloud collapse.
\end{abstract}




\section{Introduction}

Stars are formed in molecular clouds, where overdense regions (prestellar cores; \citealt{bergin2007}) are able to overcome various pressure forces and collapse under self-gravity. The processes by which these cores form and subsequently collapse are currently unclear, as the main observable quantities - dust continuum and molecular line emission - are difficult to unambiguously convert into properties such as the gas density, which can be directly compared to theoretical models. It is therefore important to determine the predictions for these quantities resulting from theory. Models of the collapse of prestellar cores can broadly be divided into those which are unstable and begin collapse immediately (e.g. \citealt{larson1969,gong2009}), and those which are magnetically subcritical (e.g. \citealt{fiedler1993}), and remain supported against collapse until the removal of magnetic support by ambipolar diffusion. The differences between these two scenarios can have important consequences for the resulting properties of the forming star: \citet{shu1987} suggested that supercritical collapse produces high-mass stars, while low-mass stars are formed from initially subcritical cores. The longer duration of collapse in subcritical models can have significant effects on the molecular abundances in the core, potentially providing a direct observational test of the process of star formation.

Previous studies, coupling by various methods the density evolution of a collapsing core with time-dependent chemistry, have confirmed that the mode of collapse can have observationally detectable consequences. \citet{aikawa2001,aikawa2003}, using the analytical solution of \citet{larson1969} and \citet{penston1969} for the dynamics, found that models with a delayed onset of collapse were in worse agreement with the observed abundances in the prestellar core L1544. \citet{aikawa2005} improved on this by post-processing hydrodynamical simulations of prestellar collapse, finding that the chemistry can be strongly affected by the initial conditions. \citet{hincelin2013,hincelin2016} performed three-dimensional magnetohydrodynamical (MHD) simulations coupled to a full chemical network in the ideal MHD limit, showing that different molecules trace different components of the core during collapse.

Non-ideal MHD, which results in effects including ambipolar diffusion, is a more complicated problem as the chemistry and dynamics are no longer independent, due to the dependence of the magnetic coupling on the ionised fraction of the gas. Many star formation studies circumvent this issue by using a reduced chemical network to calculate a large table of non-ideal MHD coefficients that are then used to interpolate the required coefficient during runtime (e.g. \citealt{tomida2015,tsukamoto2015a,tsukamoto2015b,tsukamoto2017,vaytet2018}). Although this method makes it possible to evolve the simulation for an extended period of time, the evolution of the chemical abundances are not calculated and are thus unknown at any given time. In the studies where the coefficients are calculated at runtime (e.g. \citealt{wurster2016b,wurster2017,wurster2018a,wurster2018b,wurster2018c}), only a few elements and reactions are calculated - too few to allow a meaningful comparison to observations.

To investigate the evolution of the chemical abundances, \citet{li2002} used a one-dimensional approximation to the dynamics and a simplified prescription for ion-neutral coupling, finding that their magnetised model was in better agreement with observations of L1544, seemingly in contradiction to \citet{aikawa2001,aikawa2003}. \citet{tassis2012} used a thin-disc approximation coupled to a full chemical network. They found magnetically supported models displayed significantly enhanced central depletions for many commonly observed molecules, as the longer collapse timescale allows more freeze-out of molecules onto dust grains, although the total molecular abundances, integrated over the entire core, were much less affected. \citet{tassis2012} suggested that abundance ratios between certain molecules would be a more reliable test of ambipolar diffusion, while \citet{pagani2013} found that the ortho-para H$_2$ ratio is capable of discriminating between rapid and slow collapse.

Due to the computational expense of coupling time-dependent multi-species chemistry with non-ideal MHD, previous studies have often been forced to reduce the dimensionality of the problem by various approximations. In \citet{priestley2018}, we employed an alternative approach, parametrising the results of previous (magneto)hydrodynamical simulations so that the gas density, as a function of radius and time, could be calculated from an analytical equation, which can then be trivially implemented in a time-dependent chemical simulation. We found that cores which are initally unstable against collapse show very similar chemical behaviour, despite differing initial conditions and collapse mechanisms. Our initially stable models, which collapse due to ambipolar diffusion, had noticeably different abundance profiles, generally showing depletion which was both stronger and extended to larger radii than the unstable models. However, we were unable to determine whether this was a real distinction between stable and unstable cores, or due to the differing initial conditions, in particular the density, of the simulations we parametrised. In this paper, we remove this variation by running 3D non-ideal MHD simulations of collapsing prestellar cores with mass-to-flux ratios ranging from subcritical to supercritical, and post-processing the results with a time-dependent gas-grain chemical model. As the only parameter we vary is the degree of magnetic support, this provides a clean test of potential molecular tracers of the mechanism of star formation.

\section{Method}

We simulate the collapse of a prestellar core using Phantom \citep{price2018}, an open-source smoothed-particle MHD code, combined with the NICIL library \citep{wurster2016}. NICIL self-consistently calculates the non-ideal MHD coefficients by computing the density of electrons, ions and singly-charged and neutral dust grains under local conditions, and has previously been used to study the impact of non-ideal effects on magnetic breaking and binarity in star formation \citep{wurster2016b,wurster2017}, among other topics. The initial conditions are a static, non-rotating sphere of constant density, located in a box with edge lengths $4r$ (where $r$ is the radius of the sphere). The box outside the sphere is filled with lower-density material with a density contrast of $30:1$, in pressure equilibrium with the sphere. The core is threaded by an initially uniform magnetic field aligned with the $z$ axis, although as we do not include rotation the direction of the magnetic field is essentially arbitrary. We use an isothermal equation of state, with a gas temperature of $T = 10 \kel$ and a mean molecular mass of $2.31 \mh$, corresponding to a sound speed of $c_s = 1.89 \times 10^4 \, {\rm cm} \, {\rm s}^{-1}$. The core mass is $5 \msun$, and we vary the initial radius to produce models with initial densities corresponding to $\nh \sim 10^4$ and $10^5 \pcc$, and the magnetic field strength to investigate initial mass-to-flux ratios of $5$ and $0.5$ which give magnetic field strengths covering the ranges observed in prestellar cores for the appropriate densities \citep{crutcher2010}. The input parameters for the four resulting models are listed in Table \ref{tab:hydroprop} - we refer to the $\sim 10^4 \pcc$ super- and sub-critical models as LOW-SUP and LOW-SUB, and the $\sim 10^5$ models as HIGH-SUP and HIGH-SUB, respectively. We run the supercritical models for one free-fall time, and allow the subcritical models to evolve until they have reached a comparable central density as the equivalent density supercritical model.

\begin{table*}
  \centering
  \caption{Input parameters for non-ideal MHD models.}
  \begin{tabular}{ccccccc}
    \hline
    Model & $R$ / $\pc$ & $\rho$ / g $\pcc$ & $\nh$ / $\pcc$ & $B_z$ / $\mu$G & $t_{\rm ff}$ / Myr & $t_{\rm end}$ / Myr \\
    \hline
    LOW-SUP & $0.13$ & $3.68 \times 10^{-20}$ & $1.57 \times 10^4$ & $8.08$ & $0.347$ & $0.347$  \\
    LOW-SUB & $0.13$ & $3.68 \times 10^{-20}$ & $1.57 \times 10^4$ & $80.8$ & $0.347$ & $0.972$   \\
    HIGH-SUP & $0.06$ & $3.74 \times 10^{-19}$ & $1.60 \times 10^5$ & $37.9$ & $0.109$ & $0.109$ \\
    HIGH-SUB & $0.06$ & $3.74 \times 10^{-19}$ & $1.60 \times 10^5$ & $379$ & $0.109$ & $0.239$ \\
    \hline
  \end{tabular}
  \label{tab:hydroprop}
\end{table*}

We track the chemical evolution using UCLCHEM \citep{holdship2017}, a time-dependent gas-grain chemical code incorporating freeze-out, desorption and grain-surface reactions. We select $500$ particles, evenly spaced in initial radius, from our MHD simulations, and extract the time evolution of their positions and densities. For each particle we input the time and density to a zero-dimensional (i.e. pointlike) UCLCHEM model, updating the molecular abundances at each time step, and reconstruct the three-dimensional abundance structure throughout the collapse from the particle position information. We use the UMIST12 reaction network \citep{mcelroy2013} with additional grain processes as described in \citet{holdship2017}. We assume all hydrogen is initially in molecular form, while other elements have the abundances listed in Table \ref{tab:chemprop} (the `high-metal' abundances from \citealt{lee1998}). The gas temperature is $10 \kel$ and the cosmic ray ionization rate is $1.3 \times 10^{-17} \, {\rm s}^{-1}$. We assume that the core is shielded from any external radiation fields by being embedded in a molecular cloud, and so set the radiation field intensity to zero.

\begin{table}
  \centering
  \caption{Initial abundances with respect to $\nh$.}
  \begin{tabular}{lr}
    \hline
    Quantity / unit & Value \\
    \hline
    He abundance & $0.10$ \\
    C abundance & $7.30 \times 10^{-5}$ \\
    N abundance & $2.14 \times 10^{-5}$ \\
    O abundance & $1.76 \times 10^{-4}$ \\
    S abundance & $8.00 \times 10^{-6}$ \\
    Si abundance & $8.00 \times 10^{-7}$ \\
    Mg abundance & $7.00 \times 10^{-7}$ \\
    \hline
  \end{tabular}
  \label{tab:chemprop}
\end{table}

\begin{figure}
  \centering
  \includegraphics[width=\columnwidth]{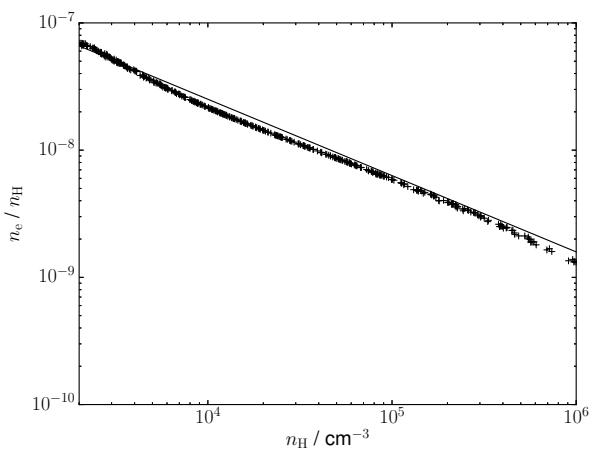}
  \caption{Electron abundance versus hydrogen nuclei density $\nh$ for the LOW-SUP model model (black crosses) and the power law function given in the text (black line).}
  \label{fig:nhne}
\end{figure}

The NICIL library calculates the ion and electron fractions of the gas assuming time equilibrium, treating cosmic ray ionization and recombination as well as processes including neutral and singly-charged dust grains. However, our chemical models include freeze-out of atoms and molecules onto grain surfaces, including Mg ions, which are the most common ionized species. The proportion of Mg locked up in grain mantles depends on the previous history of the gas, and not just the local conditions, so cannot be calculated as an equilibrium quantity. However, we find that the electron abundance follows a tight relation with the hydrogen nuclei density $\nh$, of the form ${\rm X(e^-)} = 1 \times 10^{-7} \left(\nh / 10^3 \pcc\right)^{-0.6}$, as shown in Figure \ref{fig:nhne}, so we replace the value calculated by NICIL with this function and set the ion density equal to the electron density. This ensures the values used to calculate non-ideal MHD effects and the values returned by UCLCHEM are consistent.

\section{Results}

\begin{figure}
  \centering
  \includegraphics[width=\columnwidth]{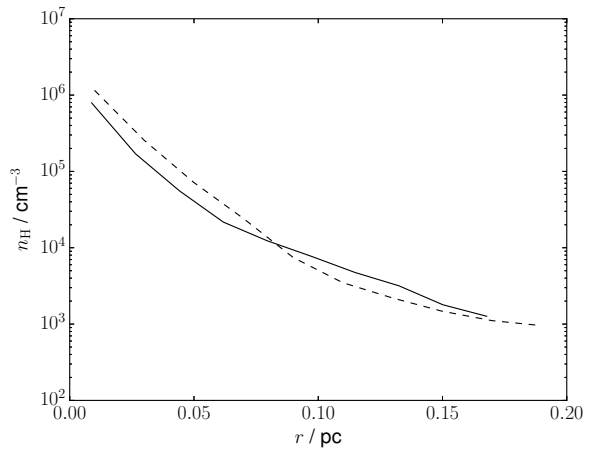}
  \caption{Hydrogen nuclei density $\nh$ in the midplane versus radius for the LOW-SUP (solid line) and LOW-SUB (dashed line) models at the end of the simulations.}
  \label{fig:dens}
\end{figure}

The supercritical models, due to the high mass-to-flux ratio, remain nearly spherically symmetric throughout the collapse, while the subcritical models retain axial symmetry. For each simulation, we bin the particles into an $r-z$ grid in cylindrical coordinates and take the average values for all particles in a bin as representative. Figure \ref{fig:dens} shows the radial density profiles in the midplane for the LOW-SUP and LOW-SUB models at the simulation end point. The radial density profiles for both models are similar, as at this point the subcritical model has contracted enough to overcome magnetic support and is collapsing dynamically, although the cloud is compressed into a thin disc rather than remaining spherical, as gas can freely collapse along the magnetic field lines.

\begin{figure}
  \centering
  \includegraphics[width=\columnwidth]{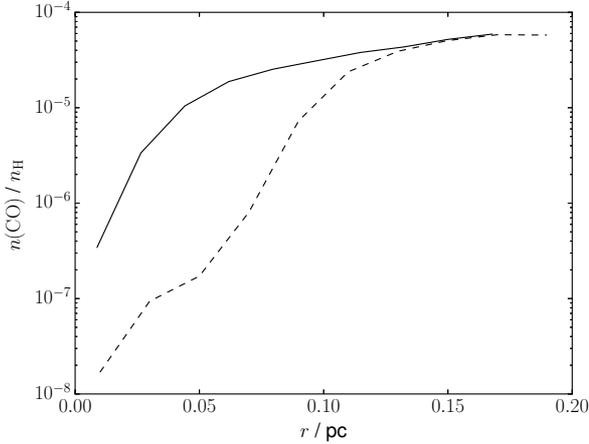}
  \caption{CO abundance in the midplane versus radius for the LOW-SUP (solid line) and LOW-SUB (dashed line) models at the end of the simulations.}
  \label{fig:coabun}
\end{figure}

Despite the similar densities, the molecular abundances differ significantly between the models. Figure \ref{fig:coabun} shows the midplane CO abundances for both models. Whereas the both models feature an abundance profile falling from a peak value of $\sim 7 \times 10^{-5}$ (the elemental abundance of carbon) at the edge to a lower, depleted value in the centre, as seen in non-magnetic models (e.g. \citealt{aikawa2005}), the subcritical model decreases in abundance much more rapidly and to more extreme values. This effect was also seen by \citet{tassis2012} and \citet{priestley2018}, and is due to the longer duration of collapse allowing depletion to proceed further in the subcritical case.

\begin{figure}
  \centering
  \includegraphics[width=\columnwidth]{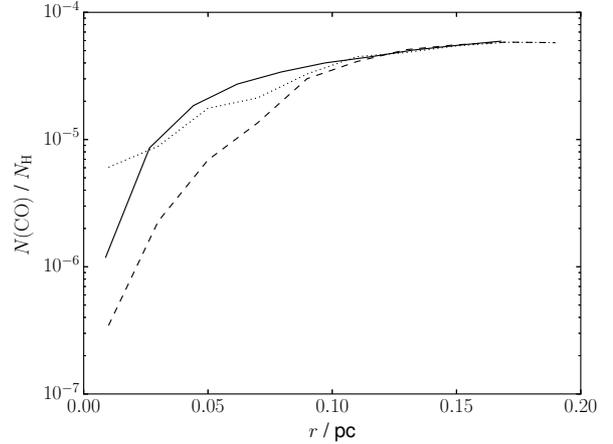}
  \caption{Integrated CO abundance versus radius for the LOW-SUP (solid line) model and { the LOW-SUB model as seen from the direction of the $z$-axis (dashed line) and edge on (dotted line)} at the end of the simulations.}
  \label{fig:cocol}
\end{figure}

While the midplane abundances are very different, observations of prestellar cores measure column, rather than volume, densities. Figure \ref{fig:cocol} shows the integrated CO abundance (weighted by density) versus radius, i.e. $N({\rm CO})/N_{\rm H}$, as seen from the direction of the $z$-axis, { and edge-on for the LOW-SUB model. We do not show the edge-on profile for the LOW-SUP model as it is essentially identical to the face on case.} The integrated abundances show much smaller differences, due to the inclusion in the LOW-SUB model of lower density, higher CO gas above and below the midplane. { Seen edge-on, the LOW-SUB profile is almost indistinguishable from the LOW-SUP case, as all sightlines pass through a significant quantity of low-density, low-depletion gas.} The total density-weighted CO abundaces for the entire core are $2.5 \times 10^{-5}$ and $1.4 \times 10^{-5}$ for the LOW-SUP and LOW-SUB models respectively. Compared to typically observed variations in the CO abundance between cores (e.g. \citealt{frau2012}), this change is too small to discriminate between the two models. The HIGH-SUP and HIGH-SUB models, with abundances of $3.3 \times 10^{-6}$ and $1.4 \times 10^{-6}$, also cannot plausibly be distinguished, although in this case the abundances are an order of magnitude lower due to the higher densities allowing freeze-out to proceed more rapidly. \citet{tassis2012} previously noted the similar total abundances of models with very different magnetic field strengths - however, we emphasise that those authors found the effect only for the integrated abundance, whereas their one-dimensional abundance profiles still show large differences depending on magnetic field strength. Our three-dimensional simulations allow us to show that even the column-averaged abundance profiles for CO remain too similar to reliably discriminate between supercritical and subcritical models using resolved observations of prestellar cores.

\begin{figure}
  \centering
  \includegraphics[width=\columnwidth]{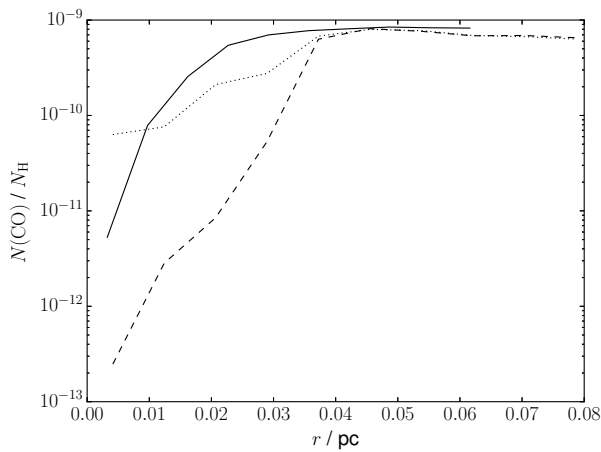}
  \caption{Integrated C$_2$H abundance versus radius for the HIGH-SUP (solid line) model and { the HIGH-SUB model as seen from the direction of the $z$-axis (dashed line) and edge on (dotted line)} at the end of the simulations.}
  \label{fig:highc2h}
\end{figure}

Table \ref{tab:abundance} lists the abundances of a number of observationally important molecules, for our models and from observations of several prestellar cores. We discuss the agreement (or otherwise) of our results with observations in Section \ref{sec:obs} - for the current discussion the observational values are presented to give an idea of the typical variation in abundances between objects. For all species, the subcritical models have lower total abundances than the supercritical model of the same density, as the longer duration results in higher depletions due to freeze-out. Similarly, the high-density models have lower abundances compared to the low-density ones - the increase in the freeze-out rate due to the increased density more than compensates for the decrease in the collapse duration. However, compared to typical variations in the observed abundances, which for some molecules are greater than a factor of ten, the differences between models of the same initial density are too small to act as a probe of the collapse rate. While differences in the inner parts of the cores can be significant for supercritical and subcritical collapse, at larger radii the abundances are generally similar. As the majority of the gas remains at these larger radii, the overall abundances for the cores are not greatly affected by differences in the inner regions. This can be seen in Figure \ref{fig:highc2h}, showing the C$_2$H integrated abundance profiles for the HIGH-SUP and HIGH-SUB models, for which the total abundance varies by a large factor ($\sim 5$) compared to most other molecules. Although C$_2$H is severely depleted out to a radius almost twice as large in the HIGH-SUB model { seen face on}, beyond this point the abundance profiles are indistinguishable, resulting in a much less dramatic difference in the total abundances. { For an edge-on orientation, as with CO the level of depletion is much lower due to gas at large radii with high abundances being in the line of sight in all cases. The central abundance is higher than the LOW-SUP model for the same reason.}

\begin{table*}
  \centering
  \caption{Total model molecular abundances and observed values from prestellar cores. Total CO abundances are derived from isotopic abundances assuming a $^{13}$C/$^{12}$C ratio of $0.011$ and an $^{18}$O/$^{16}$O ratio of $0.002$. \power{a}{b} represents $a \times 10^{b}$. Observed abundances are taken from \citet{frau2012} (Pipe Nebula), \citet{tafalla2006} (L1498, L1517B), \citet{wakelam2006} (L143N) and \citet{jorgensen2004} (L1544, L1689B).}
  \begin{tabular}{ccccccccccc}
    \hline
    Species & LOW-SUP & LOW-SUB & HIGH-SUP & HIGH-SUB & Pipe Nebula & L1498 & L1517B & L134N & L1544 & L1689B \\
    \hline
    CO & \power{2.5}{-5} & \power{1.4}{-5} & \power{3.3}{-6} & \power{1.4}{-6} & \power{0.1-2.4}{-4} & \power{2.5}{-5} & \power{7.5}{-5} & \power{8.0}{-5} & \power{4.9}{-6} & \power{2.4}{-5} \\
    CS & \power{6.4}{-8} & \power{1.7}{-8} & \power{1.2}{-8} & \power{1.0}{-8} & \power{0.2-1.5}{-9} & \power{3.0}{-9} & \power{3.0}{-9} & \power{1.7}{-9} & \power{8.9}{-10} & \power{2.6}{-9} \\
    CN & \power{1.0}{-8} & \power{5.2}{-9} & \power{1.4}{-9} & \power{3.0}{-10} & \power{0.4-2.2}{-10} & - & - & \power{8.2}{-10} & \power{4.8}{-10} & \power{<2.3}{-10} \\
    NH$_3$ & \power{3.0}{-8} & \power{2.0}{-8} & \power{4.7}{-9} & \power{2.8}{-9} & - & \power{2.8}{-8} & \power{3.4}{-8} & \power{9.1}{-8} & - & - \\
    HCN & \power{3.9}{-9} & \power{1.7}{-9} & \power{1.1}{-9} & \power{7.9}{-10} & \power{0.1-1.9}{-10} & \power{7.0}{-9} & \power{3.0}{-9} & \power{1.2}{-8} & \power{<3.5}{-10} & \power{<3.8}{-10} \\
    C$_2$H & \power{9.8}{-10} & \power{6.3}{-10} & \power{2.8}{-10} & \power{5.1}{-11} & \power{1.5-5.7}{-10} & - & - & \power{<5}{-8} & - & - \\
    C$_3$H$_2$ & \power{7.4}{-9} & \power{9.6}{-9} & \power{1.2}{-9} & \power{8.9}{-10} & \power{0.1-2.6}{-10} & \power{1.6}{-9} & \power{9.3}{-10} & \power{2.0}{-9} & - & - \\
    N$_2$H$^+$ & \power{9.5}{-11} & \power{6.1}{-11} & \power{1.6}{-11} & \power{6.0}{-12} & \power{0.5-2.7}{-11} & \power{1.7}{-10} & \power{1.5}{-10} & \power{6.8}{-10} & \power{5.0}{-9} & \power{4.3}{-10} \\
    CH$_3$OH & \power{4.7}{-10} & \power{3.1}{-10} & \power{4.9}{-11} & \power{3.4}{-11} & \power{0.2-1.4}{-9} & \power{6.0}{-10} & \power{6.0}{-10} & \power{3.7}{-9} & - & - \\
    H$_2$CO & \power{1.5}{-8} & \power{2.4}{-9} & \power{8.4}{-9} & \power{3.8}{-9} & - & \power{1.3}{-9} & \power{6.7}{-10} & \power{2.0}{-8} & - & - \\
    HCO$^+$ & \power{6.6}{-9} & \power{2.5}{-9} & \power{1.1}{-9} & \power{1.9}{-10} & - & \power{3.0}{-9} & \power{1.5}{-9} & \power{1.0}{-8} & \power{3.9}{-10} & \power{1.2}{-9} \\
    HC$_3$N & \power{7.6}{-10} & \power{1.1}{-10} & \power{6.7}{-10} & \power{9.7}{-10} & - & \power{1.0}{-9} & \power{4.5}{-10} & \power{8.7}{-10} & \power{1.6}{-9}& \power{3.6}{-11} \\
    \hline
  \end{tabular}
  \label{tab:abundance}
\end{table*}

\begin{figure}
  \centering
  \includegraphics[width=\columnwidth]{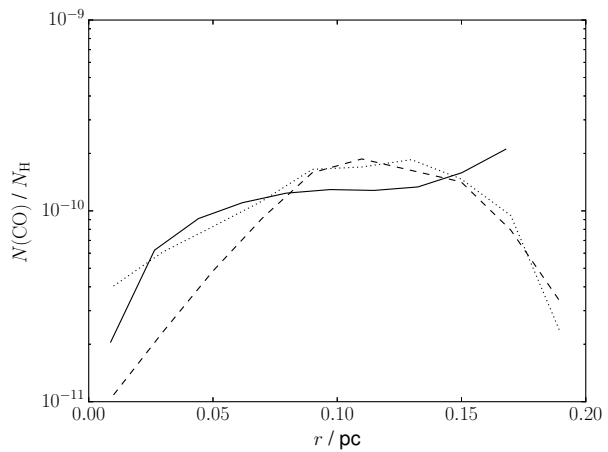}
  \caption{Integrated N$_2$H$^+$ abundance versus radius for the LOW-SUP (solid line) model and { the LOW-SUB model as seen from the direction of the $z$-axis (dashed line) and edge on (dotted line)} at the end of the simulations.}
  \label{fig:n2h+}
\end{figure}

While the magnetic field strength does not strongly affect the total abundance in the core, changes in the spatial distribution can be much more significant. As shown in Figure \ref{fig:cocol}, for CO this amounts to a decrease of a factor of a few in the LOW-SUB model compared to LOW-SUP, too small to be a reliable observational test, but differences as extreme as those of C$_2$H in the high density models (Figure \ref{fig:highc2h}) may well be detectable, { although for a core seen edge on the central abundances are much more similar}. Figure \ref{fig:n2h+} shows the N$_2$H$^+$ integrated abundance profile for the LOW-SUP and LOW-SUB models. While the abundances of the two models remain within a factor of a few, the behaviour is qualitatively different. Both models increase from the centre outward to an abundance of $\sim 10^{-10}$, but the LOW-SUP model then stays at essentially constant abundance beyond $0.05 \pc$ for a distance of $\sim 0.1 \pc$, while the LOW-SUB model peaks at $0.1 \pc$ before declining outwards. At a distance of $100 \pc$, the scales of $\sim 0.01 \pc$ required to resolve this behaviour are $\sim 20''$, easily achievable with modern radio facilities such as ALMA. Among other commonly observed molecules, we find HCO$^+$ displays similar behaviour for the low-density models, while C$_2$H, unlike in the high-density case, shows little difference.

\begin{figure}
  \centering
  \includegraphics[width=\columnwidth]{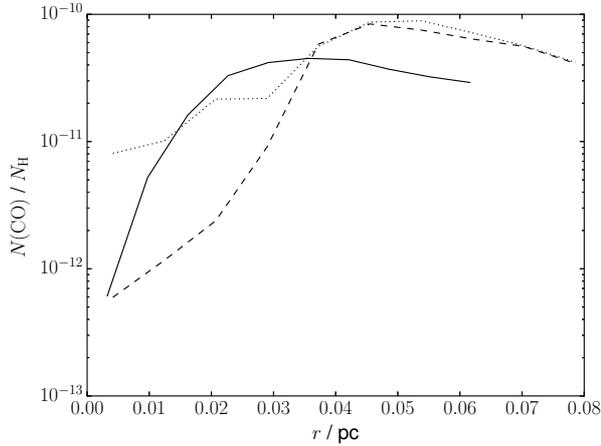}
  \caption{Integrated N$_2$H$^+$ abundance versus radius for the HIGH-SUP (solid line) model and { the HIGH-SUB model as seen from the direction of the $z$-axis (dashed line) and edge on (dotted line)} at the end of the simulations.}
  \label{fig:highn2h+}
\end{figure}

For the high-density models, the N$_2$H$^+$ and HCO$^+$ abundance profiles differ from the low-density cases, but still provide the most promising way to discriminate between supercritical and subcritical collapse. Figure \ref{fig:highn2h+} shows the N$_2$H$^+$ integrated abundance profiles - the abundance in both models increases to a essentially constant plateau, but differ in the behaviour within $\sim 0.03 \pc$ of the centre. The HIGH-SUP model increases rapidly at first before trailing off, whereas the HIGH-SUB model has a slower but accelerating increase towards the peak value, which turns over sharply into a nearly-constant value at $0.04 \pc$. { Similarly to C$_2$H, an edge-on viewing angle decreases the difference in central abundance between the models, but the difference in trend still remains.} As with the low-density models, the HCO$^+$ abundance profile behaves similarly to N$_2$H$^+$. Again, the $\sim 0.01 \pc$ resolution needed to distinguish these two scenarios is within the capabilities of current molecular observations.

A potential issue with using abundance profiles is that they require the column densities of both the molecule in question and of hydrogen to be measured. While the molecular column density can be obtained fairly simply from line intensities, as mentioned previously determining $N_{\rm H}$ can be affected by significant systematic uncertainties. With the three-dimensional structure of the core, radiative transfer modelling can be used to calculate line surface brightnesses for direct comparison with observations. We used RADEX \citep{vandertak2007} to model the emission from each $r-z$ bin, using a slab geometry, a temperature of $10 \kel$ and a linewidth of $0.5 \, {\rm km s}^{-1}$ (comparable to typical observed values; \citealt{tafalla2002,tafalla2006}), and the molecular column density calculated from each bin. We assume that the relative velocities between bins in the $z$ direction are large enough compared to the linewidths that we can ignore absorption - while this is not entirely accurate, as the velocities generally similar to our adopted linewidth value, the transitions we are interested in are not optically thick so we regard this as an acceptable approximation.

\begin{figure}
  \centering
  \includegraphics[width=\columnwidth]{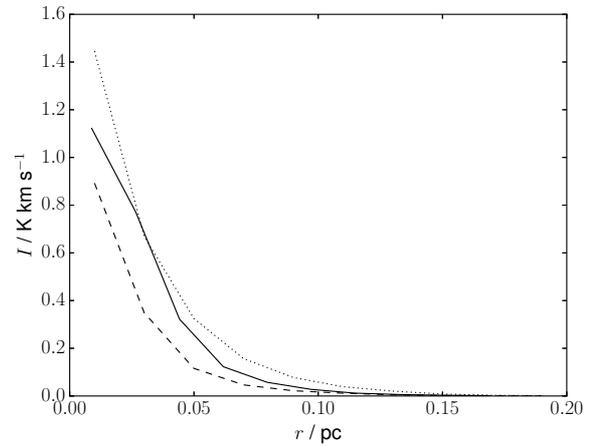}
  \caption{N$_2$H$^+$ $1-0$ line surface brightness versus radius for the LOW-SUP (solid line) model and { the LOW-SUB model as seen from the direction of the $z$-axis (dashed line) and edge on (dotted line)} at the end of the simulations.}
  \label{fig:n2h+sfb}
\end{figure}

\begin{figure}
  \centering
  \includegraphics[width=\columnwidth]{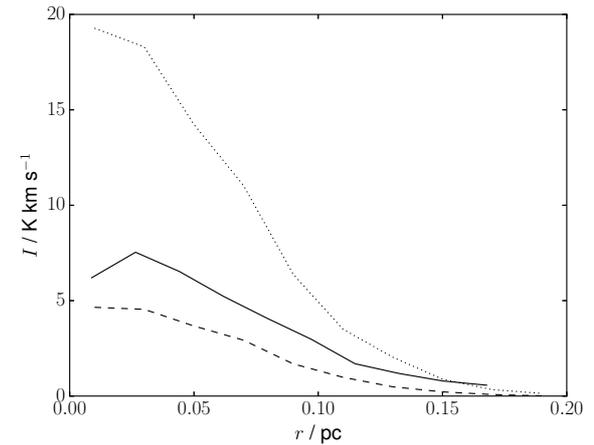}
  \caption{HCO$^+$ $1-0$ line surface brightness versus radius for the LOW-SUP (solid line) model and { the LOW-SUB model as seen from the direction of the $z$-axis (dashed line) and edge on (dotted line)} at the end of the simulations.}
  \label{fig:hco+sfb}
\end{figure}

Figure \ref{fig:n2h+sfb} shows the N$_2$H$^+$ $1-0$ line surface brightness profile for the low-density models. While the abundance profiles show notable differences, the surface brightness profiles (on a linear, rather than logarithmic scale) are similar in both shape and magnitude - compared to those in the sample of prestellar cores observed by \citet{tafalla2002}, neither model stands out as a better match. Thus for this molecule, measuring the abundance appears to be necessary in order to discriminate between collapse mechanisms. For the HCO$^+$ $1-0$ line, shown in Figure \ref{fig:hco+sfb}, the differences between the two models are greater, in particular { the edge-on LOW-SUB intensity profile, which is several times greater at the centre than the LOW-SUP case}. Compared to profiles from \citet{tafalla2006}, the LOW-SUB { face on profile} appears to be more consistent, although we note that { the surface brightnesses are} greater than those observed in the two cores in that study by a factor of several { in all cases}. The beam sizes from \citet{tafalla2006} correspond to physical sizes of $\sim 0.01 \pc$, smaller than our radial bin sizes, so beam filling effects are unlikely to be a factor.

\begin{figure}
  \centering
  \includegraphics[width=\columnwidth]{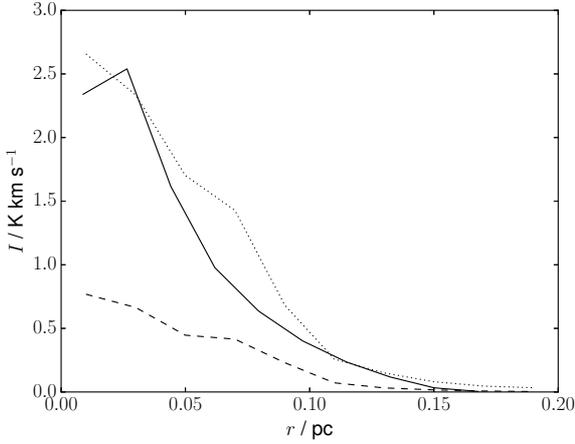}
  \caption{HCN $1-0$ line surface brightness versus radius for the LOW-SUP (solid line) model and { the LOW-SUB model as seen from the direction of the $z$-axis (dashed line) and edge on (dotted line)} at the end of the simulations.}
  \label{fig:hcnsfb}
\end{figure}

\begin{figure}
  \centering
  \includegraphics[width=\columnwidth]{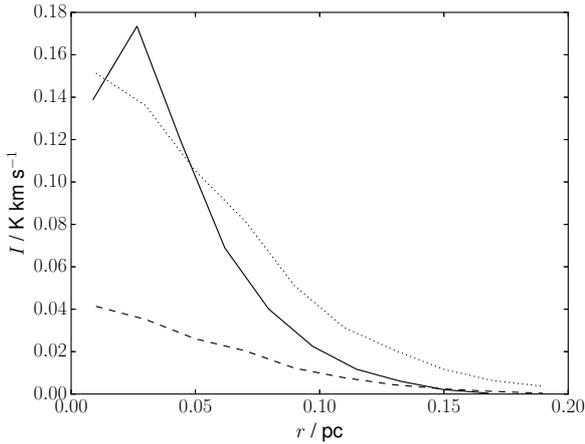}
  \caption{CH$_3$OH $2_0-1_0$ line surface brightness versus radius for the LOW-SUP (solid line) model and { the LOW-SUB model as seen from the direction of the $z$-axis (dashed line) and edge on (dotted line)} at the end of the simulations.}
  \label{fig:ch3ohsfb}
\end{figure}

While the two molecules we identify as having distinctive abundance profiles produce similar intensity profiles, the reverse is true for other typically observed species. Figures \ref{fig:hcnsfb} and \ref{fig:ch3ohsfb} show the surface brightness profiles for the HCN $1-0$ and CH$_3$OH $2_0 - 1_0$ lines, respectively, for the LOW-SUP and LOW-SUB models. In both cases, the intensities in the inner $0.05 \pc$ are higher by factors of $\sim 3$ in the LOW-SUP model { than in the LOW-SUB model seen face on}, due to the higher abundances. The LOW-SUP intensities also rise much more sharply from the outer part of the core to the centre, which seems to be incompatible with the profiles in \citet{tafalla2006} for both molecules. { Seen edge on, the LOW-SUB model produces intensity profiles more similar to the LOW-SUP case, as the line of sight passes through a greater length of large-$r$ gas where the difference between model abundances is minimal, although in this case the appearance of the core (thin disc as opposed to spherical) would allow discrimination between models.} The LOW-SUB { face on} predicted intensities are similar to those observed for HCN but lower for CH$_3$OH, whereas the LOW-SUP { and LOW-SUB edge on} values are too high for HCN but consistent with those of CH$_3$OH, although again with a different profile shape. These comparisons are not significant enough to conclusively support one model over the other, but do suggest that predicted line intensities may be a more promising test of collapse mechanisms than the molecular abundances, as the (often uncertain) measurement of $N_{\rm H}$ is no longer necessary. { In addition to the intensity, line profiles could also be used, exploiting the fact that the velocity profile differs much more than the density between supercritical and subcritical collapse. However, the number of particles the chemical evolution was followed for in this work is insufficient to produce synthetic line profiles at the required resolution (spatial and velocity), so we leave this to future work.}

\section{Discussion}
\label{sec:obs}

In order to use trends in our model abundances as a tool for investigating collapse mechanisms, our models must be at least mostly accurate representations of the chemistry in prestellar cores. Comparison with observed abundances is complicated by several factors: real cores may have had very different initial conditions from the simplistic constant-density sphere we assume, and be at various stages of collapse. The conversion from line intensity to column density and then to relative abundance introduces numerous sources of uncertainty, and the effect of beam size, which may result in the column density only being measured for a fraction of the core, can also be important. Nonetheless, if our models do represent realistic prestellar cores we would expect the abundances to broadly agree with those derived from observation.

From the abundances in Table \ref{tab:abundance}, our low-density models both predict abundances for most molecules in agreement with the observed range of values. Notable exceptions are CS and CN, and to a lesser extent C$_3$H$_2$, for which the model abundances are significantly higher than the largest observed value. N$_2$H$^+$ and HCO$^+$, which we identify as the most promising molecules for discriminating between subcritical and supercritical collapse using abundance profiles, are both consistent with observations, except for in the case of L1544 which has a higher N$_2$H$^+$ and lower HCO$^+$ abundance. L1544 is also particularly depleted in CO compared to the low-density models, and as such is probably at a later stage of evolution than we investigate - \citet{quenard2018} suggested an age of $\sim 5 \times 10^6 \yr$ for this objectbased on modelling the abundances of complex organic molecules, much later than any of our models. The abundances of HCN and CH$_3$OH, which we noted show significant differences in line intensity, are also reasonably well predicted by the models. In general, the low-density models are mostly adequate for reproducing observed chemical properties of prestellar cores.

The HIGH-SUP and HIGH-SUB models are significantly more depleted than the low-density ones, most notably with $N(CO) \sim 10^{-6}$. This is much lower than all our observational sample except for L1544 - NH$_3$ is also depleted by around an order of magnitude compared to the typical value of $\sim 10^{-8}$. The high-density models do succeed in reproducing C$_3$H$_2$ and CN, where the low-density models fail, but CS is still overabundant in the models and CH$_3$OH is much lower than observed, along with CO and NH$_3$ as mentioned previously. The excessive depletion is not just present at the end of the simulation - CO is depleted below $10^{-5}$ within a few $\times 10^{4} \yr$ in the HIGH-SUB model. Given the importance of CO both in the chemistry of prestellar cores and as an observational tracer of molecular gas, the failure to produce the observed abundances casts some doubt on whether the high-density models are a realistic description of star formation.

Comparing our results with those of \citet{tassis2012}, we find good agreement in most cases despite their models beginning from a lower density ($1000 \pcc$), treating the core as a thin disc, and directly coupling the chemical and hydrodynamical evolution of the gas. In particular, their models also predict CN abundances well above the observed values for a wide range of parameter space. Our models also predict similar values to those of \citet{aikawa2005}, who assumed the core is initially a Bonnor-Ebert sphere rather than constant density and did not include magnetic support, with the exception of CS, which can be explained by their lower sulphur abundance. This confirms our finding in \citet{priestley2018} - the initial conditions and the details of the collapse mechanism have much less effect on the chemical properties of a core than the duration of the collapse. Whereas in \citet{priestley2018} we found significant differences between subcritical and supercritical models, this appears to be due to the subcritical model evolving over $\sim 10^7 \yr$, rather than $10^6 \yr$ for the other cases we considered. In this paper, the difference between the collapse timescales of subcritical and supercritical cores is only a factor of a few, and so the resulting chemical effects are much smaller. This was also seen by \citet{tassis2012}, with the most significant change in abundances coming from models with collapse timescales $> 10 \myr$. Models where ambipolar diffusion is approximated as simply a decrease in the rate of contraction (e.g. \citealt{aikawa2001}) can therefore only be relied upon if the reduction is consistent with that expected from non-ideal MHD effects.

\section{Conclusions}

We have post-processed non-ideal MHD simulations of collapsing prestellar cores using a time-dependent gas-grain chemical model, for a range of initial densities and magnetic field strengths, allowing us to determine the three-dimensional chemical structure of the cores. Our conclusions are as follows:

\begin{itemize}
\item Magnetically supercritical and subcritical cores of the same initial density show large differences in their abundance profiles for many molecules, as the increased duration of collapse for subcritical models allows freeze-out to progress further.
\item The observable properties - abundances integrated over the column or the entire core - are less distinct, as the abundances are similar at large radii, preventing their use as a probe of the collapse mechanism.
\item The integrated abundance profiles of N$_2$H$^+$ and HCO$^+$ show qualitative differences between supercritical and subcritical models, which may allow resolved observations to distinguish between the two cases.
\item Our low-density models, with initial density $\nh = 10^4 \pcc$, are mostly consistent with observed abundances in prestellar cores. Models with initial density $10^5 \pcc$ are overdepleted in several important observational species, including CO.
\item Line surface brightness profiles, calculated from our results using radiative transfer modelling, have a greater potential for discriminating between models, as they can be compared to observations without knowledge of $N_{\rm H}$. Full three-dimensional models are required in order to convert molecular abundance information into line intensities except in highly idealised symmetric cases.
\end{itemize}

\section*{Acknowledgements}

FDP is funded by the Science and Techonology Facilities Council. JW acknowledges support from the European Research Council under the European Community's Seventh Framework Programme (FP7/2007- 2013 grant agreement no. 339248). { The authors wish to thank the referee, Konstantinos Tassis, for their helpful suggested improvements to the paper.}




\bibliographystyle{mnras}
\bibliography{ambdiff}




\bsp	
\label{lastpage}
\end{document}